\documentstyle[twocolumn,aps,prd]{revtex}
\tighten
\begin{document}
\draft
%\usepackage{amstex}   % useful for coding complex math
%\mathindent\parindent % needed in case "Amstex" is used

\title{General-covariant evolution formalism
for Numerical Relativity}
\author{C.~Bona, T.~Ledvinka$^{\dag}$, C.~Palenzuela
       and M.~\v Z\' a\v cek$^{\dag}$}

\address{
Departament de Fisica, Universitat de les Illes Balears,
Ctra de Valldemossa km 7.5, 07071 Palma de Mallorca, Spain\\
$^{\dag}$Institute of Theoretical Physics, Faculty of Mathematics
and Physics, Charles University, V Hole\v{s}ovi\v{c}k\'ach 2, 180
00 Prague 8, Czech Republic}
\maketitle

\begin{abstract}
A general covariant extension of Einstein\'{}s field equations is
considered with a view to Numerical Relativity applications. The
basic variables are taken to be the metric tensor and an
additional four-vector $Z_\mu$. Einstein's solutions are recovered
when the additional four-vector vanishes, so that the energy and
momentum constraints amount to the covariant algebraic condition
$Z_\mu=0$. The extended field equations can be supplemented by
suitable coordinate conditions in order to provide symmetric
hyperbolic evolution systems: this is actually the case for either
harmonic coordinates or normal coordinates with harmonic slicing.
\end{abstract}

\section{Introduction}

General covariance is a key feature of General Relativity. At a
first look, Einstein\'{}s field equations can be understood as a
set of ten second order partial differential equations on the ten
unknown metric coefficients $g_{\mu\nu}$
\begin{equation}\label{Einstein}
  R_{\mu \nu} = 8\; \pi\; (T_{\mu \nu} - \frac{1}{2} \;T \; g_{\mu \nu})
\end{equation}
However, in General Relativity, like in many other field theories,
physical solutions are given by an equivalence class of field
values, related to one another by gauge transformations. General
covariance means that the gauge group in Einstein\'{}s theory is
that of the general (smooth) coordinate transformations
\begin{equation}\label{4covariance}
  y^{\mu} = f^{\mu}(x^{\nu}).
\end{equation}
Accordingly, the field equations (\ref{Einstein}) do not (even may
not) provide enough information to determine the values of the ten
unknown coefficients $g_{\mu\nu}$. On the other hand, in numerical
applications we must deal with specific metric components. It
follows that a numerical evolution system must include a
specification of the coordinates as an extra ingredient in order
to determine the four kinematical degrees of freedom. Fixing four
of the ten metric coefficients we choose one specific expression
for $g_{\mu\nu}$ out of the equivalence class representing the
same physical solution. A general covariant evolution system would
be incomplete and conversely, a complete evolution system can not
preserve general covariance.

To be more specific, we will carefully distinguish evolution
systems, associated with a particular gauge choice, from evolution
formalisms. The later can be defined as a set of equations that
apply to an equivalence class of solutions, prior to a complete
gauge specification. Following this way, we will be able to draw a
line between general covariant formalisms, where the equivalence
class is defined by the full group of coordinate transformations
(\ref{4covariance}) and partially covariant formalisms, where the
equivalence class is defined by any restricted subset of
coordinate transformations. This means also for instance that
different evolution systems can be obtained from the same
evolution formalism, as we will see in the following paragraphs.

\subsection{General Covariant Formalisms}
A good example of a general covariant formalism is provided by
recent numerical relativity works \cite{SW02,Garf02} based upon a
well known classical approach. The evolution formalism is given by
the original field equations (\ref{Einstein}), although the
DeDonder \cite{DeDo21,Fock59} expression of the Ricci tensor is
used to write down the principal part, namely
\begin{equation}\label{DeDonder}
 -\Box\; g_{\mu \nu} + \partial_{\mu} \Gamma_{\nu}
  + \partial_{\nu} \Gamma_{\mu}  = ...
\end{equation}
where the box symbol stands for the d\'{}Alembert operator on
functions and we have noted $\Gamma^{\mu} \equiv g^{\rho \sigma}
{\Gamma^{\mu}}_{\rho \sigma}$ as usual. General covariance is not
lost in passing from (\ref{Einstein}) to (\ref{DeDonder}), because
we only reordered the partial derivatives in the Ricci tensor.

Is is obvious from a comparison between (\ref{DeDonder}) and the
wave equation for $g_{\mu\nu}$ that we can obtain a symmetric
hyperbolic system imposing the well known harmonic coordinate
conditions \cite{DeDo21,Fock59}:
\begin{equation}\label{full_harmonic}
 \Box\; x^{\mu} = - \Gamma^{\mu} = 0.
\end{equation}
Although the first proofs of well-posedness of the resulting
evolution system were well known \cite{Choq55,FM72}, the
corresponding proof for the initial-boundary problem, which is
highly relevant for Numerical Relativity applications, has been
given recently \cite{SW02,FN99,CPSTR02}. Different evolution
systems can be obtained from the general covariant formalism
(\ref{DeDonder}) by modifying the harmonic coordinate conditions
(\ref{full_harmonic}) \cite{HE73,DeTu81}. One can even add
arbitrary "gauge source" terms to the right-hand-side of
(\ref{full_harmonic}) to obtain a wide class of generalized
harmonic evolution systems \cite{Frie85}.

A different evolution formalism can be obtained by using the well
known 3+1 decomposition \cite{Choq56,ADM62}, where one considers
the space-time sliced by $t=constant$ hypersurfaces. The line
element can be written as
\begin{equation}\label{line3+1}
    ds^2 = - \alpha^2\;dt^2  
    + \gamma_{ij}\;(dx^i+\beta^i\;dt)\;(dx^j+\beta^j\;dt),
\end{equation}
where the lapse $\alpha$ and the shift $\beta^i$ represent the
kinematical degrees of freedom. The field equations
(\ref{Einstein}) can then be translated in terms of the
three-dimensional geometry of the slices, namely
\begin{eqnarray}
  &&(\partial_t -{\cal L}_{\beta}) \gamma_{ij} = - 2\; \alpha\;
  K_{ij}
\label{K} \\
  &&(\partial_t -{\cal L}_{\beta}) K_{ij} = -\nabla_i\alpha_j
    + \alpha\;  [{}^{(3)}R_{ij}-2K^2_{ij}+trK\;K_{ij}]
\label{evolve_K_Einstein} \\
  &&^{(3)}R - tr(K^2) + (trK)^2 = 0
\label{energy_constraint} \\
  &&\nabla_k ({K^k}_i - {\delta^k}_i\; trK) = 0
\label{momentum_constraint}
\end{eqnarray}
where we have restricted ourselves to the vacuum case for
simplicity.

Let us notice that the coordinate gauge freedom is not limited in
any way by translating the four-dimensional (4D) field equations
(\ref{Einstein}) into the 3+1 version
(\ref{K}-\ref{momentum_constraint}). The lapse $\alpha$ and the
shift $\beta^i$ can take arbitrary values, so that the four gauge
degrees of freedom are still at our disposal. Although general
covariance is not manifest in the 3+1 equations
(\ref{K}-\ref{momentum_constraint}), their solution space is still
invariant under general coordinate transformations
(\ref{4covariance}), because it is equivalent to the corresponding
solution space of the 4D equations (\ref{Einstein}). Conversely,
in the 3+1 version (\ref{K}-\ref{momentum_constraint}) it is
manifestly clear that one can evolve the dynamical degrees of
freedom $\gamma_{ij} (t,x^k)$ from any consistent set of initial
data $\{ \gamma_{ij} (0,x^k), K_{ij} (0,x^k)  \}$ using the
evolution equations (\ref{K},\ref{evolve_K_Einstein}). The
remaining ones (\ref{energy_constraint},\ref{momentum_constraint})
can be interpreted as constraints. This diversity in the evolution
properties of both sets of equations is not obvious in the 4D
version. To summarize: both the 4D equations (\ref{Einstein}) and
the 3+1 equations provide equivalent general-covariant formalisms,
although general covariance is apparent only in the 4D version,
whereas the evolution properties are manifest only in the 3+1
version (\ref{K}-\ref{momentum_constraint}).

\subsection{Extending solution space: taking constraints out}

The 3+1 formalism (\ref{K}-\ref{momentum_constraint}) is specially
suited for Numerical Relativity applications. In this context, one
usually takes advantage of the fact that the energy and momentum
constraints (\ref{energy_constraint},\ref{momentum_constraint})
are first integrals of (\ref{K},\ref{evolve_K_Einstein}). This
allows us to enforce the constraints
(\ref{energy_constraint},\ref{momentum_constraint}) on the initial
and boundary data only, or even to use them to monitor the
accuracy of the time evolution. But the constraints are not
enforced by the time evolution algorithm for interior points,
which in its simplest form is based only on
(\ref{K},\ref{evolve_K_Einstein}) (free evolution approach
\cite{Cent80,Naka80}). This "unconstrained" evolution formalism,
although perfectly consistent \cite{Chop91,Frit97}, does introduce
a strong discrimination between the two sets of equations
(\ref{K},\ref{evolve_K_Einstein}) and
(\ref{energy_constraint},\ref{momentum_constraint}) that breaks
the general covariance of the 3+1 formalism.

To verify this, let us notice first that by replacing the full 3+1
formalism (\ref{K}-\ref{momentum_constraint}) by the subset
(\ref{K},\ref{evolve_K_Einstein}) we are actually extending the
solution space so as to include constraint-violating pairs $\{
\gamma_{ij}, K_{ij}\}$. Now, as the restricted set of equations
(\ref{K},\ref{evolve_K_Einstein}) corresponds to the space
components of (\ref{Einstein}), the extended space of solutions
will be invariant only under the restricted subset of coordinate
transformations (\ref{4covariance}) that preserve the time
slicing, namely
\begin{eqnarray}\label{3covariance}
    t'   &=& h(t) \\
    y^{i} &=& f^i(x^j,t). \nonumber
\end{eqnarray}
This confirms that general covariance is broken in the
unconstrained evolution formalisms.

Let us remark that the extension of the solution space of
(\ref{K}-\ref{momentum_constraint}) is a rule, not an exception,
among the new formalisms arising after the seminal 1983 work of
Y.~Choquet-Bruhat and T.~Ruggeri \cite{CR83}, which opened the
door to the use of arbitrary shift choices in hyperbolic evolution
systems. The bottom line is that the constraints
(\ref{energy_constraint},\ref{momentum_constraint}) contained in
the original 3+1 evolution formalism are at odds with
hyperbolicity to the intent that some extension is needed in order
to modify the mathematical structure of the formalism without
loosing the physical solutions. The easiest way of doing this is
just taking the constraints
(\ref{energy_constraint},\ref{momentum_constraint}) out of the
system. This is the basic ingredient, although this crucial point
can be masked by other manipulations on the evolution equations,
like taking an extra time derivative \cite{CR83,AACY95}, or an
extra space derivative \cite{Frie96}, or using the constraints to
modify the evolution equations (\ref{K},\ref{evolve_K_Einstein})
\cite{FR94,FR96,AY99,Hern00,KST01,ST02}. The resulting formalisms,
when supplemented with suitable coordinate conditions, provide
hyperbolic evolution systems that can be used in Numerical Relativity
applications. From our point of view, these formalisms can also be
interpreted as providing many non-equivalent ways of extending the
solution space of (\ref{K}-\ref{momentum_constraint}) with at
least two related common features: constraint equations
(\ref{energy_constraint},\ref{momentum_constraint}) are left out
of the final evolution formalism and general covariance is broken
as a result, even before a specific coordinate system is selected.

\subsection{Extending solution space: extra dynamical fields}

A completely different way of extending the solution space is to
introduce extra dynamical fields, independent of the metric and
its derivatives, into the evolution formalism. This alternative
has been independently used by many Numerical Relativity groups in
different ways \cite{BM92,BMSS95,SN95,BS99}. The key idea in these
works was to introduce three supplementary dynamical fields whose
evolution equations were obtained by using the momentum constraint
(\ref{momentum_constraint}). As far as these works were focused on
Numerical Relativity applications
\cite{YS02,BB02,ABDKPST02}, the
supplementary quantities were introduced in an "ad hoc" way,
breaking even the 3+1 covariance (\ref{3covariance}) of the
formalism. Only very recently \cite{BLP02} the same idea has been
implemented in a way which is at least invariant under the
restricted subset of coordinate trasformations
(\ref{3covariance}): the extra quantities are given by a
three-dimensional "zero" vector $Z_i$ which vanishes for
Einstein\'{}s solutions. During numerical evolution, however,
non-zero values of $Z_i$ arise due to truncation errors and the
resulting numerical codes actually deal with an extended set of
solutions.

Even with this improvement, general covariance is still broken for
two different reasons. First of all, although the momentum
constraint (\ref{momentum_constraint}) has been incorporated into
the formalism as the right-hand-side of the time evolution
equation for $Z_i$, the energy constraint
(\ref{energy_constraint}) is still taken out of the time evolution
algorithm. It is obvious that the extension of solution space to
"energy constraint violating" modes can not be invariant under the
general coordinate transformations (\ref{4covariance}). This
reason alone could be easily overcome by proceeding along the
lines sketched in \cite{BFHR99}, where every constraint is
incorporated into the system by adding an extra "Lagrange
multiplier" quantity: every extra quantity could then be coupled
with the previous ones in many different ways so that many more
new arbitrary parameters will appear in the resulting evolution
formalism. But this would just reinforce the second cause of
general covariance breaking: the lack of covariance of the final
set of extra quantities. If one wants to extend solution space by
adding extra fields without breaking general covariance, then this
set of extra fields should be equivalent to some set of well
defined space-time quantities, independent of the time slicing
considered.

\section{General covariant extended evolution formalisms}

\subsection{The extended field equations}

We propose to extend the field equations (\ref{Einstein}) in a
general covariant way by introducing an extra four-vector $Z_\mu$,
so that the set of basic fields will consist into the pair $\{
g_{\mu\nu}, Z_\mu \}$. The original field equations
(\ref{Einstein}) will then be replaced by
\begin{equation}\label{Einstein_extended}
  R_{\mu \nu} + \nabla_{\mu} Z_{\nu} + \nabla_{\nu} Z_{\mu} =
  8\; \pi\; (T_{\mu \nu} - \frac{1}{2}\;T\; g_{\mu \nu}).
\end{equation}
The solutions of the Einstein\'{}s solutions can be easily
recognized among the extended set as those satisfying condition
\cite{nota1}
\begin{equation}\label{Zeq0}
  Z_\mu = 0
\end{equation}
so that the four-vector $Z_\mu$ will provide a simple way to
monitor the quality of numerical simulations or any other kind of
approximation scheme. Notice that equations
(\ref{Einstein_extended}) are of mixed order: second order in the
metric components $g_{\mu\nu}$, but only first order in the extra
vector field $Z_\mu$ . This means in particular that terms
containing first derivatives of $Z_\mu$  belong to the principal
part and that they are then relevant to the causal structure of
the resulting evolution systems, as we will see later.

In order to fully understand the evolution properties of the
extended equations, let us translate the manifestly covariant form
(\ref{Einstein_extended}) into the 3+1 language (\ref{line3+1}).
The covariant four-vector $Z_\mu$ will then be decomposed into its
space components $Z_i$ and the normal component
\begin{equation}\label{theta}
    \Theta \equiv n_{\mu}\; Z^{\mu} = \alpha\; Z^0
\end{equation}
where $n_\mu$ is the unit normal to the $t=constant$ slices. The
4D equations (\ref{Einstein_extended}) can then be written in the
equivalent form:
\begin{eqnarray}
   (\partial_t - {\cal L}_{\beta}) \gamma_{ij} &=& -2\; \alpha\;
   K_{ij}
\label{evolve_metric_ext} \\
   (\partial_t - {\cal L}_{\beta}) K_{ij} &=& -\nabla_i\alpha_j
   + \alpha\; [{}^{(3)}R_{ij} + \nabla_i Z_j + \nabla_j Z_i
\label{evolve_K_ext} \\
   &-& 2\;K^2_{ij}+ (trK - 2\; \Theta)\;K_{ij}]
\nonumber\\
 (\partial_t -{\cal L}_{\beta}) \Theta &=& \frac{\alpha}{2}\;
 [{}^{(3)}R + (trK - 2\; \Theta)\;trK
\label{evolve_theta_ext}  \\
 &-& tr(K^2) + 2\; \nabla_k Z^k  - 2\; ({\alpha}_k/\alpha)\; Z^k]
\nonumber \\
 (\partial_t -{\cal L}_{\beta}) Z_i &=& \alpha\;
 [\nabla_k\;({K_i}^k -{\delta_i}^k trK) + \partial_i \Theta
\label{evolve_Z_ext} \\
   &-& ({\alpha}_i/\alpha)\; \Theta  -2\; {K_i}^k\; Z_k]
\nonumber
\end{eqnarray}
where $\alpha_i$ stands for $\partial_i \alpha$ and we have
restricted ourselves to the vacuum case again.

The evolution properties of the formalism are transparent in the
3+1 version (\ref{evolve_metric_ext}-\ref{evolve_Z_ext}). Only
evolution equations appear there, without constraints. One can not
omit any of the equations from
(\ref{evolve_metric_ext}-\ref{evolve_Z_ext}) because all them are
needed to evolve the extended set of ten dynamical fields $\{
\gamma_{ij},\Theta,Z_i \}$. This means that general covariance
will not be broken in Numerical Relativity applications because
all the equations (\ref{evolve_metric_ext}-\ref{evolve_Z_ext})
must be used on equal footing in the main algorithm evolving the
dynamical fields in time: there is no room left for equation
discrimination.

\subsection{Recovering Einstein\'s solutions}

The algebraic condition (\ref{Zeq0}) is useful to check \emph{a
posteriori} whether a given solution of the extended system
(\ref{Einstein_extended}) is actually a solution of the Einstein's
field equations (\ref{Einstein}). But it is interesting as well to
know \emph{a priori} the necessary and sufficient condition for a
given set of initial data to generate a physical solution. In this
sense, one can take the divergence of the extended equations
(\ref{Einstein_extended}) to get, allowing for the contracted
Bianchi identities,
\begin{equation}\label{laplacian}
  \Box Z_{\mu} + R_{\mu \nu}\; Z^{\nu} = 0.
\end{equation}
This homogeneous second order equation in $Z_\mu$ ensures that any
deviation from the original Einstein equations (\ref{Einstein})
propagates through light cones and also that a sufficient set of
conditions for the initial data to provide physical solutions is
given by
\begin{equation}\label{hom_conditions}
   Z_{\mu}(0,x^i) = 0 \:\:\:\:\:\: \partial_t Z_{\mu}(0,x^i) = 0,
\end{equation}
where the second equation, allowing for (\ref{evolve_theta_ext},
\ref{evolve_Z_ext}), represents imposing energy and momentum
constraints (\ref{energy_constraint},\ref{momentum_constraint}) on
the initial data.

This means that the algebraic constraint (\ref{Zeq0}) by itself is
not a first integral of the extended system
(\ref{Einstein_extended}). Equations (\ref{energy_constraint},
\ref{momentum_constraint}) appear here as auxiliary conditions so
that the full set (\ref{energy_constraint},
\ref{momentum_constraint}, \ref{Zeq0}) is preserved by time
evolution. From the practical point of view, this means that one
can take any set of consistent initial data of Einstein's
equations (\ref{Einstein}) and use it with a zero initial value of
$Z_\mu$ to get an initial data set for the extended equations
(\ref{Einstein_extended}) that will generate precisely the same
solution. The general-covariant gauge-independent equation
(\ref{laplacian}) can be then interpreted as a useful tool to
understand the propagation of the initial constraints
(\ref{hom_conditions}).

\section{Coordinate conditions and symmetric hyperbolic evolution systems}

As stated in the introduction, the evolution system is not
complete until one provides coordinate conditions to fix the four
kinematical degrees of freedom. We will consider here two
different coordinate conditions leading to a symmetric hyperbolic
evolution system. We will start in both cases from one of the two
equivalent versions of the extended general-covariant formalism
(\ref{Einstein_extended}).

\subsection{A 4D evolution system in harmonic coordinates}

Let us use the DeDonder \cite{DeDo21,Fock59} expression of the
Ricci tensor to write down the principal part of
(\ref{Einstein_extended}), namely
\begin{equation}\label{DeDonder_ext}
 -\Box\; g_{\mu \nu} + \partial_{\mu} (\Gamma_{\nu} + 2\; Z_{\nu})
   + \partial_{\nu} (\Gamma_{\mu} + 2\; Z_{\mu}) = ...
\end{equation}
It is obvious from a comparison with the wave equation for
$g_{\mu\nu}$ that we can obtain a symmetric hyperbolic system
\begin{equation}\label{symmetric_metric}
 \Box\; g_{\mu\nu} = ...
\end{equation}
provided that we kill the additional terms in (\ref{DeDonder_ext})
using the following extension of the well known harmonic
coordinate conditions:
\begin{equation}\label{full_harmonic_ext}
 \Box\; x^{\mu} = - \Gamma^{\mu} = 2\; Z^{\mu}\:\:\:.
\end{equation}
The four-vector $Z_\mu$ can be interpreted in this context as a
sort of "gauge source", along the lines sketched in ref.
\cite{Frie85}.

Notice that, in contrast with the classical approach, the
compatibility between the reduced system (\ref{symmetric_metric})
and the coordinate conditions (\ref{full_harmonic_ext}) is not an
issue in the present context because there are now fourteen
independent components of the fields $\{g_{\mu\nu}, Z_\mu \}$ to
be determined by the fourteen equations (\ref{symmetric_metric},
\ref{full_harmonic_ext}). A straightforward analysis in the 3+1
framework shows that this is actually the case: the lapse and
shift evolution is provided by equation (\ref{full_harmonic_ext})
and the evolution of the remaining ten degrees of freedom,
including the four-vector $Z_\mu$, is given by
(\ref{symmetric_metric}). Notice also that equation
(\ref{full_harmonic_ext}) actually coincides with the classical
harmonic coordinate condition (\ref{full_harmonic}) for physical
solutions, where $Z_\mu$ vanishes. This is why we talk about
"harmonic coordinates" to refer also to condition
(\ref{full_harmonic_ext}) in the present context.

\subsection{A 3+1 evolution system with harmonic slicing}

Harmonic coordinates are not flexible enough to be used in most
Numerical Relativity applications. A more suitable choice is the
"harmonic slicing", in which the time coordinate is again assumed
to be a harmonic function, but the space coordinates are chosen so
that the mixed components $g_{0i}$ vanish (normal coordinates).
We propose here to keep in the extended case the time component of
(\ref{full_harmonic_ext}), that is
\begin{equation}\label{harmonic_slicing}
  g_{0i}=0 \:,\:\:\:\: \Box\; x^{0} = - \Gamma^{0} = 2\; Z^{0}
\end{equation}
which can be translated into the 3+1 language as
\begin{equation}\label{harmonic_slicing_3+1}
  \beta^{i}=0 \:,\:\:\:\:
  \partial_t  \ln \alpha = - \alpha\; (trK-2\; \Theta).
\end{equation}
The principal part of (\ref{evolve_K_ext}-\ref{evolve_Z_ext}) can
be written as
\begin{eqnarray}
   &\partial_t& K_{ij} + \partial_k [\alpha\;{\lambda^k}_{ij}] = ...
\label{evolve_K_harm} \\
  &\partial_t& Z_i +
   \partial_k [\alpha({\delta^k}_i\; (trK - \Theta) - {K^k}_i)]  = ...
\label{evolve_Z_harm} \\
   &\partial_t& \Theta + \partial_k [\alpha\;(D^k - E^k - Z^k)] = ...
\label{evolve_Theta_harm}
\end{eqnarray}
where we have noted
\begin{eqnarray}
    &&A_k \equiv \partial_k (\ln \alpha), \:\:\:\:
    D_{kij} \equiv \frac{1}{2}\; \partial_k \gamma_{ij}
\label{A_D} \\
   &&{\lambda^k}_{ij} = {D^k}_{ij} +
    \frac{{\delta^k}_i}{2}(A_j+D_j-2\;E_j-2\;Z_j)
\label{flux_K} \\
    && \:\:\:\:\:\:\:\:\:\:\:
    +\frac{{\delta^k}_j}{2}(A_i+D_i-2\;E_i-2\;Z_i)
\end{eqnarray}
and $D_k =\gamma^{ij}\;D_{kij}$, $E_k =\gamma^{ij}\;D_{ijk}$. 

One can get a fully first order system in the usual way, by considering
both $A_k$ and $D_{kij}$ as independent additional quantities. Their
evolution equations can be then easily obtained by differentiating
(\ref{evolve_metric_ext}) and (\ref{harmonic_slicing_3+1}), namely
\begin{eqnarray}
    &&\partial_t A_k + \partial_k [\alpha (trK - 2\;\Theta)] = 0
\label{evolve_A} \\
   &&\partial_t D_{kij} + \partial_k [\alpha\; K_{ij}] = 0
\label{evolve_D}
\end{eqnarray}
The full set of basic independent quantities is then given by
$\{\alpha,\gamma_{ij},K_{ij},\Theta,Z_i,A_k,D_{kij} \}$ and the
non-trivial principal part of the corresponding evolution systems
is given by equations (\ref{evolve_K_harm}-\ref{evolve_Theta_harm},
\ref{evolve_A},\ref{evolve_D}).

The causal structure of the first order system
(\ref{harmonic_slicing_3+1}-\ref{evolve_D}) is simpler
than expected: one can easily check from either
(\ref{DeDonder_ext},\ref{harmonic_slicing}) or
(\ref{harmonic_slicing_3+1}-\ref{evolve_D}) that
\begin{equation}\label{evolve_mixed}
 \partial_t (\Gamma_i + 2\; Z_i) =
 \partial_t (2\;E_i - D_i - A_i + 2\; Z_i)= ...
\end{equation}
(the quantities $\Gamma_i$ here are just the space components of the
quantities $\Gamma_\mu$ in eq. \ref{DeDonder_ext}) so that one readily identifies 
three eigenfields propagating along the normal lines. These
"standing modes" are the only deviation of our system with respect
to the wave equation pattern: all the remaining non-trivial eigenfields
propagate along light cones. If we select a specific space direction,
along a given unit vector $u_k$, then these light cone eigenfields are
\begin{equation}\label{eigenvectors}
    K_{ij} \pm u_k {\lambda^k}_{ij} \;\;,\;\;\;
    (trK -2\;\Theta) \pm u_k A^k.
\end{equation}
A straightforward calculation shows then that the first order
system (\ref{harmonic_slicing_3+1}-\ref{evolve_Theta_harm}) is
symmetric hyperbolic. The "symmetrizer" can be easily identified
starting from quadratic positive-definite "Energy" functions which
are conserved up to lower order terms. One such "Energy estimates"
is provided by
\begin{eqnarray}
  E &=& K^{ij}K_{ij} + \lambda^{kij}\lambda_{kij} +
  (trK - 2\;\Theta)^2
\label{energy} \\
   &+& A^k A_k + (\Gamma^k+2\;Z^k)(\Gamma_k+2\;Z_k).
\nonumber
\end{eqnarray}
We are currently working with finite difference algorithms that
can take advantage of the symmetric hyperbolicity of the system
(\ref{harmonic_slicing_3+1}-\ref{evolve_D}) to increase the
robustness of numerical simulations. Our preliminary results
indicate that explicit "Energy" expressions of the form
(\ref{energy}) can be very useful to devise stable boundary
conditions, along the lines sketched in refs \cite{CPSTR02,LS02}.

{\em Acknowledgements: The authors are indebted with A.Arbona,
J.Mass\'o and M.Tiglio for useful discussions and comments.
This work has been supported by the EU Programme 'Improving the
Human Research Potential and the Socio-Economic Knowledge Base'
(Research Training Network Contract (HPRN-CT-2000-00137), by the
Spanish Ministerio de Ciencia y Tecnologia through the research
grant number BFM2001-0988 and by a grant from the Conselleria
d'Innovacio i Energia of the Govern de les Illes Balears.}

\bibliographystyle{prsty}

\begin{thebibliography}{99}

% IBVP for Einstein equations with harmonic gauge
\bibitem{SW02} B.~Szil\'{a}gyi and J.~Winicour,
              gr-qc/0205044.

\bibitem{Garf02} D.~Garfinkle, Phys.~Rev.~D{\bf 65} 044029 (2002).

% decomposition of Einstein into Dedonder
\bibitem{DeDo21} T.~De.~Donder,{\em La Gravifique Einstenienne}
       Gauthier-Villars, Paris (1921).

% decomposition of Einstein into Dedonder
\bibitem{Fock59} Fock, V.A., {\em The theory of Space Time and Gravitation},
       Pergamon, London (1959).

% proof that the IV problem "Einstein+harmonic coordinates" is well possed
\bibitem{Choq55} Y.~Choquet-Bruhat, Acta Math. {\bf 88} 141
           (1955).

% proof that the IV problem "Einstein+harmonic coordinates" is well possed
%Symmetric hyperbolic evolution in harmonic coordinates in GR
\bibitem{FM72} A.~Fischer and J.~Marsden, Commun.~Math.~Phys. {\bf 28},
         1-28 (1972).

%introduce the maximally dissipative technique into GR
\bibitem{FN99} H.~Friedrich and G.~Nagy, Commun.~Math.~Phys. {\bf 201},
         619 (1999).

% proof that the IBVP for the linerized Einstein equations is well possed
\bibitem{CPSTR02} G.~Calabrese, J.~Pullin, O.~Sarbach, M.~Tiglio and
        O.~Reula, gr-qc/0209017.

\bibitem{HE73} S.~W.~Hawking and G.~F.~R.~Ellis, {\em The large
         scale structure of spacetime}, Cambridge U.P. (1973).

\bibitem{DeTu81} D.~DeTurck, Invent.~Math. {\bf 65} 179 (1981).

\bibitem{Frie85} H.~Friedrich, Commun.~Math.~Phys. {\bf 100},
         525 (1985).

% 3+1 decomposition
\bibitem{Choq56} Y.~Choquet-Bruhat, J.~Rat.~Mec.~Analysis {\bf 5},
         951 (1956).

% 3+1 decomposition
\bibitem{ADM62} R.~Arnowit, S.~Deser and C.~W.~Misner, {\em Gravitation:
        an introduction to current research}, ed. L.~Witten,
        Wiley, New York (1962).

\bibitem{Cent80} J.~Centrella, Phys.~Rev.~D {\bf 21}, 2776 (1980).

\bibitem{Naka80} T.~Nakamura, Prog.~Theor.~Phys. {\bf 65}, 1876 (1980).

% consistent algorithm for Einstein equations
\bibitem{Chop91} M.~W.~Choptuik, Phys.~Rev.~D {\bf 44}, 3124 (1991).

% note on the propagation of the constraints
\bibitem{Frit97} S.~Frittelli, Phys.~Rev.~D {\bf 55}, 5992 (1997).

\bibitem{CR83} Y.~Choquet-Bruhat and T.~Ruggeri,
         Comm.~Math.~Phys. {\bf 89}, 269 (1983).

\bibitem{AACY95} A.~Abrahams, A.~Anderson, Y.~Choquet-Bruhat and
        J.~W.~York,  Phys.~Rev.~Lett. {\bf 75}, 3377 (1995).

% hyperbolic reductions for Einstein equations
\bibitem{Frie96} H.~Friedrich, Class.~Quantum.~Grav. {\bf 13}, 1451 (1996).

% symmetric hyperbolic system
\bibitem{FR94} S.~Frittelli and O.~A.~Reula,
        Commun.~Math.~Phys. {\bf 166}, 221 (1994).

% symmetric hyperbolic system
\bibitem{FR96} S.~Frittelli and O.~A.~Reula,
        Phys.~Rev.~Lett. {\bf 76}, 4667 (1996).

% symmetric hyperbolic system (EC)
\bibitem{AY99} A.~Anderson and J.~W.~York, Jr.,
        Phys.~Rev.~Lett. {\bf 82}, 4384 (1999).

% symmetric hyperbolic system (adding constraints to EC)
\bibitem{Hern00} S.~D.~Hern, Ph.~D. Thesis, gr-qc/0004036.

% symmetric hyperbolic system (adding constraings ans change
% of variables to EC)
\bibitem{KST01} L.~E.~Kidder, M.~A.~Scheel and
         S.~A.~Teukolsky, Phys.~Rev. {\bf D64}, 064017 (2001).

% symmetric hyperbolic systems with live gauge
\bibitem{ST02} O.~Sarbach and M.~Tiglio,  Phys.~Rev. {\bf D66},
        064023 (2002).

% first hyperbolic system in FOFCH form
\bibitem{BM92} C.~Bona and J.~Mass\'o,
        Phys.~Rev.~Lett. {\bf 68}, 1097 (1992)

% BM system with Vs...the best one of BM
\bibitem{BMSS95} C.~Bona, J.~Mass\'o, E.~Seidel and J. Stela,
        Phys.Rev.Lett. {\bf 75}, 600 (1995)

% BSSN system, the first part
\bibitem{SN95} M.~Shibata and T.~Nakamura,
        Phys.~Rev.~D{\bf 52}, 5428 (1995).

% BSSN system, the second part
\bibitem{BS99} T.~W.~Baumgarte and S.~L.~Shapiro,
        Phys.~Rev.~D {\bf 59}, 024007 (1999).

\bibitem{YS02} G.~Yoneda and H.~Shinkai, gr-qc/0204002 (2002).

% numerical test of evolution systems
\bibitem{BB02} J.~M.~Bardeen and L.~T.~Buchman, 
         Phys.~Rev.~D {\bf 65}, 064037 (2002).

\bibitem{ABDKPST02} M.~Alcubierre, B.~Bruegmann, P.~Diener, M.~ Koppitz,
         D.~Pollney, E.~Seidel and R.~Takahashi, gr-qc/0206072 (2002).

% Z3 system, comparing with BSSN,BM and KST
\bibitem{BLP02} C.~Bona, T.~Ledvinka, C.~Palenzuela,
        Phys.~Rev.~D {\bf 66}, 084013 (2002).

\bibitem{nota1} This sufficient condition is actually the only
alternative in generic spacetimes, without any non-trivial Killing
vector field.

% Lambda system, with assymptotically stable constraint propagation
\bibitem{BFHR99} O.~Brodbeck, S.~Fritelli, P.~Hubner and O.~Reula,
        J.~Math.~Phys. {\bf 40}, 909 (1999).

%energy norms and stability of the EEE;interesting initial data
\bibitem{LS02} L.~Lindblom and M.~Scheel,
         Phys.~Rev.~D {\bf 66}, 084014  (2002).

\end{thebibliography}

\end{document}